% PLAIN TEX!

 % MACROS

%
% TEXT
%

% fonts

\def\famname{
 \textfont0=\textrm \scriptfont0=\scriptrm
 \scriptscriptfont0=\sscriptrm
 \textfont1=\textmi \scriptfont1=\scriptmi
 \scriptscriptfont1=\sscriptmi
 \textfont2=\textsy \scriptfont2=\scriptsy \scriptscriptfont2=\sscriptsy
 \textfont3=\textex \scriptfont3=\textex \scriptscriptfont3=\textex
 \textfont4=\textbf \scriptfont4=\scriptbf \scriptscriptfont4=\sscriptbf
 \skewchar\textmi='177 \skewchar\scriptmi='177
 \skewchar\sscriptmi='177
 \skewchar\textsy='60 \skewchar\scriptsy='60
 \skewchar\sscriptsy='60
 \def\rm{\fam0 \textrm} \def\bf{\fam4 \textbf}}
\def\sca#1{scaled\magstep#1} \def\scah{scaled\magstephalf} 
\def\twelvepoint{
 \font\textrm=cmr12 \font\scriptrm=cmr8 \font\sscriptrm=cmr6
 \font\textmi=cmmi12 \font\scriptmi=cmmi8 \font\sscriptmi=cmmi6 
 \font\textsy=cmsy10 \sca1 \font\scriptsy=cmsy8
 \font\sscriptsy=cmsy6
 \font\textex=cmex10 \sca1
 \font\textbf=cmbx12 \font\scriptbf=cmbx8 \font\sscriptbf=cmbx6
 \font\it=cmti12
 \font\sectfont=cmbx12 \sca1
 \font\sectmath=cmmib10 \sca2
 \font\sectsymb=cmbsy10 \sca2
 \font\refrm=cmr10 \scah \font\refit=cmti10 \scah
 \font\refbf=cmbx10 \scah
 \def\twelverm{\textrm} \def\twelveit{\it} \def\twelvebf{\textbf}
 \famname \textrm 
 \advance\voffset by .06in \advance\hoffset by .28in
 \normalbaselineskip=17.5pt plus 1pt \baselineskip=\normalbaselineskip
 \parindent=21pt
 \setbox\strutbox=\hbox{\vrule height10.5pt depth4pt width0pt}}

% primes and other "@"garbage

\catcode`@=11

{\catcode`\'=\active \def'{{}^\bgroup\prim@s}}

\def\screwcount{\alloc@0\count\countdef\insc@unt}   % for stupid
\def\screwdimen{\alloc@1\dimen\dimendef\insc@unt} % \outer errors
\def\screwbox{\alloc@4\box\chardef\insc@unt}

\catcode`@=12

% Text style parameters ("textile"?)

\overfullrule=0pt			% gets rid of stupid black boxes
\vsize=9in \hsize=6in
%\parskip=\medskipamount	% space between paragraphs
\lineskip=0pt				% minimum box separation
\abovedisplayskip=1.2em plus.3em minus.9em % space above equation
\belowdisplayskip=1.2em plus.3em minus.9em	% " below
\abovedisplayshortskip=0em plus.3em	% " above when no overlap
\belowdisplayshortskip=.7em plus.3em minus.4em	% " below
\parindent=21pt
\setbox\strutbox=\hbox{\vrule height10.5pt depth4pt width0pt}
\def\makefootline{\baselineskip=30pt \line{\the\footline}}
\footline={\ifnum\count0=1 \hfil \else\hss\twelverm\folio\hss \fi}
\pageno=1

% Box at relative coordinates (x,y)

\def\put(#1,#2)#3{\screwdimen\unit  \unit=1in
	\vbox to0pt{\kern-#2\unit\hbox{\kern#1\unit
	\vbox{#3}}\vss}\nointerlineskip}

% Lines & stuff

\def\\{\hfil\break}
\def\newpage{\vfill\eject}
\def\center{\leftskip=0pt plus 1fill \rightskip=\leftskip \parindent=0pt
 \def\textindent##1{\par\hangindent21pt\footrm\noindent\hskip21pt
 \llap{##1\enspace}\ignorespaces}\par}
 % use as {\center ... \par}, shorten lines with \\
\def\unnarrower{\leftskip=0pt \rightskip=\leftskip}

%% Page and section headings and reference stuff (also, \refs below)

%\def\sect#1\par{\par\ifdim\lastskip<\medskipamount
%	\bigskip\medskip\goodbreak\else\nobreak\fi
%	\noindent{\sectfont{#1}}\par\nobreak\medskip} % see Ü below 
%\def\itemize#1 {\item{[#1]}}	% see £ below; also use \itemitem 
\def\vol#1 {{\refbf#1} }		 % see É below
%\def\topic{\par\noindent \hangafter1 \hangindent20pt}
%\def\topic{\par\noindent \hangafter1 \hangindent60pt}
%\def\ref#1{\setbox0=\hbox{M}$\vbox to\ht0{}^{#1}$}

% Journal abbreviations

\def\NP #1 {{\refit Nucl. Phys.} {\refbf B{#1}} }
\def\PL #1 {{\refit Phys. Lett.} {\refbf{#1}} }
\def\PR #1 {{\refit Phys. Rev. Lett.} {\refbf{#1}} }
\def\PRD #1 {{\refit Phys. Rev.} {\refbf D{#1}} }

% More nitpicking

\hyphenation{pre-print}
\hyphenation{quan-ti-za-tion}

%
% MATH (mostly)
%

% accent over:

\def\oonoo#1#2#3{\vbox{\ialign{##\crcr
	\hfil\hfil\hfil{$#3{#1}$}\hfil\crcr\noalign{\kern1pt\nointerlineskip}
	$#3{#2}$\crcr}}}
\def\oon#1#2{\mathchoice{\oonoo{#1}{#2}{\displaystyle}}
	{\oonoo{#1}{#2}{\textstyle}}{\oonoo{#1}{#2}{\scriptstyle}}
	{\oonoo{#1}{#2}{\scriptscriptstyle}}}
\def\dt#1{\oon{\hbox{\bf .}}{#1}}  
\def\ddt#1{\oon{\hbox{\bf .\kern-1pt.}}#1}    % À À   (see below)
\def\slap#1#2{\setbox0=\hbox{$#1{#2}$}
	#2\kern-\wd0{\hfuzz=1pt\hbox to\wd0{\hfil$#1{/}$\hfil}}}
\def\sla#1{\mathpalette\slap{#1}}                % slash: see Ö   below
\def\bop#1{\setbox0=\hbox{$#1M$}\mkern1.5mu
	\lower.02\ht0\vbox{\hrule height0pt depth.06\ht0
	\hbox{\vrule width.06\ht0 height.9\ht0 \kern.9\ht0
	\vrule width.06\ht0}\hrule height.06\ht0}\mkern1.5mu}
\def\bo{{\mathpalette\bop{}}}                        % box: see õ below
\def~{\widetilde} % tilde key; use Option-N for accent in text & math,
	% Option-' for sim, Option-0 for math space (¼)
\mathcode`\*="702A                  % * now always complex conjugate
\def\in{\relax\ifmmode\mathchar"3232\else{\refit in\/}\fi} % ã below 
	   % fraction
\def\half{{\textstyle{1\over{\raise.1ex\hbox{$\scriptstyle{2}$}}}}}

% stick with math italic for cap Greek
\def\Gamma{\mathchar"0100}
\def\Delta{\mathchar"0101}
\def\Theta{\mathchar"0102}
\def\Lambda{\mathchar"0103}
\def\Xi{\mathchar"0104}
\def\Pi{\mathchar"0105}
\def\Sigma{\mathchar"0106}
\def\Upsilon{\mathchar"0107}
\def\Phi{\mathchar"0108}
\def\Psi{\mathchar"0109}
\def\Omega{\mathchar"010A}

\catcode128=13 \def €{\"A}                 % Option-u A
\catcode129=13 \def {\AA}                 % Option-A
\catcode130=13 \def '{\c}           	   % Option-C (cedilla)
\catcode131=13 \def ƒ{\'E}                   % Option-e E
\catcode132=13 \def "{\~N}                   % Option-n N
\catcode133=13 \def …{\"O}                 % Option-u O
\catcode134=13 \def †{\"U}                  % Option-u U
\catcode135=13 \def ‡{\'a}                  % Option-e a
\catcode136=13 \def ˆ{\`a}                   % Option-`  a
\catcode137=13 \def ‰{\^a}                 % Option-i a
\catcode138=13 \def Š{\"a}                 % Option-u a
\catcode139=13 \def ‹{\~a}                   % Option-n a
\catcode140=13 \def Œ{\alpha}            % Option-a
\catcode141=13 \def {\chi}                % Option-c
\catcode142=13 \def Ž{\'e}                   % Option-e e
\catcode143=13 \def {\`e}                    % Option-`  e
\catcode144=13 \def {\^e}                  % Option-i e
\catcode145=13 \def '{\"e}                % Option-u e
\catcode146=13 \def '{\'\i}                 % Option-e i
\catcode147=13 \def "{\`\i}                  % Option-`  i
\catcode148=13 \def "{\^\i}                % Option-i i
\catcode149=13 \def •{\"\i}                % Option-u i
\catcode150=13 \def –{\~n}                  % Option-n n
\catcode151=13 \def —{\'o}                 % Option-e o
\catcode152=13 \def ˜{\`o}                  % Option-`  o
\catcode153=13 \def ™{\^o}                % Option-i o
\catcode154=13 \def š{\"o}                 % Option-u o
\catcode155=13 \def ›{\~o}                  % Option-n o
\catcode156=13 \def œ{\'u}                  % Option-e u
\catcode157=13 \def {\`u}                  % Option-`  u
\catcode158=13 \def ž{\^u}                % Option-i u
\catcode159=13 \def Ÿ{\"u}                % Option-u u
\catcode160=13 \def  {\tau}               % Option-t
\catcode161=13 \mathchardef ¡="2203     % Option-* (TeX's usual eq. *)
\catcode162=13 \def ¢{\oplus}           % Option-4
\catcode163=13 \def £{\relax\ifmmode\to\else\itemize\fi} % Option-3
\catcode164=13 \def ¤{\subset}	  % Option-6
\catcode165=13 \def ¥{\infty}           % Option-8
\catcode166=13 \def ¦{\mp}                % Option-7
\catcode167=13 \def §{\sigma}           % Option-s
\catcode168=13 \def ¨{\rho}               % Option-r
\catcode169=13 \def ©{\gamma}         % Option-g
\catcode170=13 \def ª{\leftrightarrow} % Option-2 ; Option-E (acute) :
\catcode171=13 \def «{\relax\ifmmode\acute\else\expandafter\'\fi}
\catcode172=13 \def ¬{\relax\ifmmode\expandafter\ddt\else\expandafter\"\fi}
\catcode173=13 \def ­{\equiv}            % Option-= ; ^ Option-U (umlaudt)
\catcode174=13 \def ®{\approx}          % Option-"
\catcode175=13 \def ¯{\Omega}          % Option-O
\catcode176=13 \def °{\otimes}          % Option-5
\catcode177=13 \def ±{\ne}                 % Option-+
\catcode178=13 \def ²{\le}                   % Option-,
\catcode179=13 \def ³{\ge}                  % Option-.
\catcode180=13 \def ´{\upsilon}          % Option-y
\catcode181=13 \def µ{\mu}                % Option-m
\catcode182=13 \def ¶{\delta}             % Option-d
\catcode183=13 \def ·{\epsilon}          % Option-w
\catcode184=13 \def ¸{\Pi}                  % Option-P
\catcode185=13 \def ¹{\pi}                  % Option-p
\catcode186=13 \def º{\beta}               % Option-b
\catcode187=13 \def »{\partial}           % Option-9
\catcode188=13 \def ¼{\nobreak\ }       % Option-0
\catcode189=13 \def ½{\zeta}               % Option-z
\catcode190=13 \def ¾{\sim}                 % Option-'
\catcode191=13 \def ¿{\omega}           % Option-o
\catcode192=13 \def À{\dt}                     % Option-?
\catcode193=13 \def Á{\gets}                % Option-1
\catcode194=13 \def Â{\lambda}           % Option-l
\catcode195=13 \def Ã{\nu}                   % Option-v
\catcode196=13 \def Ä{\phi}                  % Option-f
\catcode197=13 \def Å{\xi}                     % Option-x
\catcode198=13 \def Æ{\psi}                  % Option-j
\catcode199=13 \def Ç{\int}                    % Option-\
\catcode200=13 \def È{\oint}                 % Option-|
\catcode201=13 \def É{\relax\ifmmode\cdot\else\vol\fi}    % Option-;
\catcode202=13 \def Ê{\relax\ifmmode\,\else\thinspace\fi}
\catcode203=13 \def Ë{\`A}                      % Option-`  A ; ^ Option-space
\catcode204=13 \def Ì{\~A}                      % Option-n A
\catcode205=13 \def Í{\~O}                      % Option-n O
\catcode206=13 \def Î{\Theta}              % Option-Q
\catcode207=13 \def Ï{\theta}               % Option-q; Option-- :
\catcode208=13 \def Ð{\relax\ifmmode\bar\else\expandafter\=\fi}
\catcode209=13 \def Ñ{\overline}             % Option-_
\catcode210=13 \def Ò{\langle}               % Option-[
\catcode211=13 \def Ó{\relax\ifmmode\{\else\ital\fi}      % Option-{
\catcode212=13 \def Ô{\rangle}               % Option-]
\catcode213=13 \def Õ{\}}                        % Option-}
\catcode214=13 \def Ö{\sla}                      % Option-/; Option-V :
\catcode215=13 \def ×{\relax\ifmmode\check\else\expandafter\v\fi}
\catcode216=13 \def Ø{\"y}                     % Option-u y
\catcode217=13 \def Ù{\"Y}  		    % Option-u Y
\catcode218=13 \def Ú{\Leftarrow}       % Option-!
\catcode219=13 \def Û{\Leftrightarrow}       % Option-@ ; Option-# :
\catcode220=13 \def Ü{\relax\ifmmode\Rightarrow\else\sect\fi}
\catcode221=13 \def Ý{\sum}                  % Option-$
\catcode222=13 \def Þ{\prod}                 % Option-%
\catcode223=13 \def ß{\widehat}              % Option-^
\catcode224=13 \def à{\pm}                     % Option-&
\catcode225=13 \def á{\nabla}                % Option-(
\catcode226=13 \def â{\quad}                 % Option-)
\catcode227=13 \def ã{\in}               	% Option-W
\catcode228=13 \def ä{\star}      	      % Option-R
\catcode229=13 \def å{\sqrt}                   % Option-M
\catcode230=13 \def æ{\^E}			% Option-i E
\catcode231=13 \def ç{\Upsilon}              % Option-Y
\catcode232=13 \def è{\"E}    	   	 % Option-u E
\catcode233=13 \def é{\`E}               	  % Option-`  E
\catcode234=13 \def ê{\Sigma}                % Option-S
\catcode235=13 \def ë{\Delta}                 % Option-D
\catcode236=13 \def ì{\Phi}                     % Option-F
\catcode237=13 \def í{\`I}        		   % Option-`  I
\catcode238=13 \def î{\iota}        	     % Option-H
\catcode239=13 \def ï{\Psi}                     % Option-J
\catcode240=13 \def ð{\times}                  % Option-K
\catcode241=13 \def ñ{\Lambda}             % Option-L
\catcode242=13 \def ò{\cdots}                % Option-:
\catcode243=13 \def ó{\^U}			% Option-i U
\catcode244=13 \def ô{\`U}    	              % Option-`  U
\catcode245=13 \def õ{\bo}                       % Option-B ; Option-I :
\catcode246=13 \def ö{\relax\ifmmode\hat\else\expandafter\^\fi}
\catcode247=13 \def÷{\relax\ifmmode\tilde\else\expandafter\~\fi}
\catcode248=13 \def ø{\ll}                         % Option-< ; ^ Option-N
\catcode249=13 \def ù{\gg}                       % Option-> 
\catcode250=13 \def ú{\eta}                      % Option-h 
\catcode251=13 \def û{\kappa}                  % Option-k 
\catcode252=13 \def ü{\half}     		 % Option-Z 
\catcode253=13 \def ý{\Gamma} 		% Option-G 
\catcode254=13 \def þ{\Xi}   			% Option-X ; Option-T : 
\catcode255=13 \def ÿ{\relax\ifmmode{}^{\dagger}{}\else\dag\fi}

% hat, check, tilde, and bar have been defined to work in text as well.

\def\ital#1Õ{{\it#1\/}}	     % for italics in text: see Ó above
\def\un#1{\relax\ifmmode\underline#1\else $\underline{\hbox{#1}}$
	\relax\fi}

	% for unitalicized
\def\roonoo#1#2#3{\vbox{\ialign{##\crcr
	\hfil{$#3{#1}$}\hfil\crcr\noalign{\kern1pt\nointerlineskip}
	$#3{#2}$\crcr}}}

	% accent under
\def\tdt#1{\oon{\hbox{\bf .\kern-1pt.\kern-1pt.}}#1}   % À À À
\def\({\eqno(}

%\def\refs{\sect{REFERENCES}\par\medskip \frenchspacing 
%	\parskip=0pt \refrm \baselineskip=1.23em plus 1pt
%	\def\ital##1Õ{{\refit##1\/}}}

% Young tableaux:  \upõ<a>{\õ<a>...\õ<b>}

\def\õ#1{
	\screwcount\num
	\num=1
	\screwdimen\downsy
	\downsy=-1.5ex
	\mkern-3.5mu
	õ
	\loop
	\ifnum\num<#1
	\llap{\raise\num\downsy\hbox{$õ$}}
	\advance\num by1
	\repeat}
\def\upõ#1#2{\screwcount\numup
	\numup=#1
	\advance\numup by-1
	\screwdimen\upsy
	\upsy=.75ex
	\mkern3.5mu
	\raise\numup\upsy\hbox{$#2$}}

%%%%%%%%%%%%%%%%%%%%%%%%%%%%%%%%%%%%%%%%

% postscript/pdf

\newcount\marknumber	\marknumber=1
\newcount\countdp \newcount\countwd \newcount\countht 

%
% for ordinary tex
%
\ifx\pdfoutput\undefined
\def\rgboo#1{}
\input epsf

\def\postscript#1{\special{" #1}}		% for dvips
\postscript{
	/bd {bind def} bind def
	/fsd {findfont exch scalefont def} bd
	/sms {setfont moveto show} bd
	/ms {moveto show} bd
	/pdfmark where		% printers ignore pdfmarks
	{pop} {userdict /pdfmark /cleartomark load put} ifelse
	[ /PageMode /UseOutlines		% bookmark window open
	/DOCVIEW pdfmark}
\def\bookmark#1#2{\postscript{		% #1=subheadings (if not 0)
	[ /Dest /MyDest\the\marknumber /View [ /XYZ null null null ] /DEST pdfmark
	[ /Title (#2) /Count #1 /Dest /MyDest\the\marknumber /OUT pdfmark}%
	\advance\marknumber by1}
\def\pdfklink#1#2{%
	\hskip-.25em\setbox0=\hbox{#1}%
		\countdp=\dp0 \countwd=\wd0 \countht=\ht0%
		\divide\countdp by65536 \divide\countwd by65536%
			\divide\countht by65536%
		\advance\countdp by1 \advance\countwd by1%
			\advance\countht by1%
		\def\linkdp{\the\countdp} \def\linkwd{\the\countwd}%
			\def\linkht{\the\countht}%
	\postscript{
		[ /Rect [ -1.5 -\linkdp.0 0\linkwd.0 0\linkht.5 ] 
		/Border [ 0 0 0 ]
		/Action << /Subtype /URI /URI (#2) >>
		/Subtype /Link
		/ANN pdfmark}{\rgb{1 0 0}{#1}}}
%
% for pdftex
%
\else
\def\rgboo#1{\pdfliteral{#1 rg #1 RG}}

\pdfcatalog{/PageMode /UseOutlines}		% bookmark window open
\def\bookmark#1#2{
	\pdfdest num \marknumber xyz
	\pdfoutline goto num \marknumber count #1 {#2}
	\advance\marknumber by1}
\def\pdfklink#1#2{%
	\noindent\pdfstartlink user
		{/Subtype /Link
		/Border [ 0 0 0 ]
		/A << /S /URI /URI (#2) >>}{\rgb{1 0 0}{#1}}%
	\pdfendlink}
\fi

\def\rgbo#1#2{\rgboo{#1}#2\rgboo{0 0 0}}
\def\rgb#1#2{\mark{#1}\rgbo{#1}{#2}\mark{0 0 0}}
\def\pdflink#1{\pdfklink{#1}{#1}}
\def\xxxlink#1{\pdfklink{[arXiv:#1]}{http://arXiv.org/abs/#1}}

\catcode`@=11

\def\wlog#1{}	% I don't care about new registers

% headers/footers

\def\makeheadline{\vbox to\z@{\vskip-36.5\p@
	\line{\vbox to8.5\p@{}\the\headline%
	\ifnum\pageno=\z@\rgboo{0 0 0}\else\rgboo{\topmark}\fi%
	}\vss}\nointerlineskip}
\headline={
	\ifnum\pageno=\z@
		\hfil
	\else
		\ifnum\pageno<\z@
			\ifodd\pageno
				\tenrm\romannumeral-\pageno\hfil\lefthead\hfil
			\else
				\tenrm\hfil\righthead\hfil\romannumeral-\pageno
			\fi
		\else
			\ifodd\pageno
				\tenrm\hfil\righthead\hfil\number\pageno
			\else
				\tenrm\number\pageno\hfil\lefthead\hfil
			\fi
		\fi
	\fi}

\catcode`@=12

\def\righthead{\hfil} \def\lefthead{\hfil}
\nopagenumbers

% divisions

%\def\bulletfill{\cleaders\hbox{$\mathsurround=0pt \mkern4mu
%	\raise.15em\hbox{$\bullet$} \mkern4mu$}\hfill}
\def\chrulefill{\rgb{1 0 0}{\hrulefill}}
\def\cdotfill{\rgb{1 0 0}{\dotfill}}
\newcount\area	\area=1
\newcount\cross	\cross=1
\def\volume#1\par{\newpage\noindent{\biggest{\rgb{1 .5 0}{#1}}}
	\par\nobreak\bigskip\medskip\area=0}
\def\chapskip{\par\ifnum\area=0\bigskip\medskip\goodbreak
	\else\newpage\fi}
\def\chapy#1{\area=1\cross=0
	\xdef\lefthead{\rgbo{1 0 .5}{#1}}\vbox{\biggerer\offinterlineskip
	\line{\chrulefill¼\hphantom{\lefthead}\chrulefill}
	\line{\chrulefill¼\lefthead\chrulefill}}\par\nobreak\medskip}
\def\chap#1\par{\chapskip\bookmark3{#1}\chapy{#1}}
\def\sectskip{\par\ifnum\cross=0\bigskip\medskip\goodbreak
	\else\newpage\fi}
\def\secty#1{\cross=1
	\xdef\righthead{\rgbo{1 0 1}{#1}}\vbox{\bigger\offinterlineskip
	\line{\cdotfill¼\hphantom{\righthead}\cdotfill}
	\line{\cdotfill¼\righthead\cdotfill}}\par\nobreak\medskip}
\def\sect#1 #2\par{\sectskip\bookmark{#1}{#2}\secty{#2}}
\def\subsectskip{\par\ifdim\lastskip<\medskipamount
	\bigskip\medskip\goodbreak\else\nobreak\fi}
\def\subsecty#1{\noindent{\sectfont{\rgbo{.5 0 1}{#1}}}\par\nobreak\medskip}
\def\subsect#1\par{\subsectskip\bookmark0{#1}\subsecty{#1}}
\long\def\x#1 #2\par{\hangindent2\parindent%
\mark{0 0 1}\rgboo{0 0 1}{\bf Exercise #1}\\#2%
\par\rgboo{0 0 0}\mark{0 0 0}}
\def\refs{\bigskip\noindent{\bf \rgbo{0 .5 1}{REFERENCES}}\par\nobreak\medskip
	\frenchspacing \parskip=0pt \refrm \baselineskip=1.23em plus 1pt
	\def\ital##1Õ{{\refit##1\/}}}
\long\def\twocolumn#1#2{\hbox to\hsize{\vtop{\hsize=2.9in#1}
	\hfil\vtop{\hsize=2.9in #2}}}

% fonts

\twelvepoint
\font\bigger=cmbx12 \sca2
\font\biggerer=cmb10 \sca5
%\font\biggest=cmssdc10 scaled 3583
\font\biggest=cmssdc10 scaled 4500
%\font\subtitlefont=cmbxti10 scaled 3583
 \sca5

 \sca3

% symbols

\def Ü{\relax\ifmmode\Rightarrow\else\expandafter\subsect\fi}
\def Û{\relax\ifmmode\Leftrightarrow\else\expandafter\sect\fi}
\def Ú{\relax\ifmmode\Leftarrow\else\expandafter\chap\fi}

\def\itemize#1 {\item{\bf#1}}
\def\itemizze#1 {\itemitem{\bf#1}}
\def\itemutem{\par\indent\indent \hangindent3\parindent \textindent}
\def\itemizzze#1 {\itemutem{\bf#1}}
\def ª{\relax\ifmmode\leftrightarrow\else\itemizze\fi}
\def Á{\relax\ifmmode\gets\else\itemizzze\fi}

\def\¢{\ominus}

\def\Ä{\varphi}  \def\¿{\varpi}	\def\Ï{\vartheta}

\def ò{\relax\ifmmode\cdots\else\dotfill\fi}

% boxes drawn around phrases or paragraphs

\def\cvrule{\rgbo{0 .5 1}{\vrule}}
\def\chrule{\rgbo{0 .5 1}{\hrule}}
\def\boxit#1{\leavevmode\thinspace\hbox{\cvrule\vtop{\vbox{\chrule%
	\vskip3pt\kern1pt\hbox{\vphantom{\bf/}\thinspace\thinspace%
	{\bf#1}\thinspace\thinspace}}\kern1pt\vskip3pt\chrule}\cvrule}%
	\thinspace}
\def\Boxit#1{\noindent\vbox{\chrule\hbox{\cvrule\kern3pt\vbox{
	\advance\hsize-7pt\vskip-\parskip\kern3pt\bf#1
	\hbox{\vrule height0pt depth\dp\strutbox width0pt}
	\kern3pt}\kern3pt\cvrule}\chrule}}

% boxes around equations

          % inside $$'s
   % outside $$'s

% other

\def\today{\ifcase\month\or
 January\or February\or March\or April\or May\or June\or July\or
 August\or September\or October\or November\or December\fi
 \space\number\day, \number\year}

\parindent=20pt
\newskip\normalparskip	\normalparskip=.7\medskipamount
\parskip=\normalparskip	% space between paragraphs

%%%%%%%%%%%%%%%%%%%%%%%%%%%%%%%%%%%%%%%%

% Some stupid little things that have to go at the end:
% make |,<,> OK in text

\catcode`\|=\active \catcode`\<=\active \catcode`\>=\active 
\def|{\relax\ifmmode\delimiter"026A30C \else$\mathchar"026A$\fi}
\def<{\relax\ifmmode\mathchar"313C \else$\mathchar"313C$\fi}
\def>{\relax\ifmmode\mathchar"313E \else$\mathchar"313E$\fi}

%%%%%%%%%%%%%%%%%%%%%%%%%%%%%%%%%%%%%%%%

% PAPER:
%	\paper
%
%	"title"
%
%	"authors"
%
%	"preprint number"
%
%	"date"
%
%	"abstract"
%
%	"text"
%
%	\bye

\def\thetitle#1#2#3#4#5{
 \def\titlefont{\biggest} \font\footrm=cmr10 \font\footit=cmti10
  \twelverm
	{\hbox to\hsize{#4 \hfill YITP-SB-#3}}\par
	\vskip.8in minus.1in {\center\baselineskip=2.2\normalbaselineskip
 {\titlefont #1}\par}{\center\baselineskip=\normalbaselineskip
 \vskip.5in minus.2in #2
	\vskip1.4in minus1.2in {\twelvebf ABSTRACT}\par}
 \vskip.1in\par
 \narrower\par#5\par\unnarrower\vskip3.5in minus3.3in\eject}
\def\paper\par#1\par#2\par#3\par#4\par#5\par{
	\thetitle{#1}{#2}{#3}{#4}{#5}} 
\def\author#1#2{#1 \vskip.1in {\twelveit #2}\vskip.1in}
\def\YITP{C. N. Yang Institute for Theoretical Physics\\
	State University of New York, Stony Brook, NY 11794-3840}
\def\WS{W. Siegel\footnote{$*$}{% e.g.,\author\WS\YITP
	\pdflink{mailto:siegel@insti.physics.sunysb.edu}\\
	\pdfklink{http://insti.physics.sunysb.edu/\~{}siegel/plan.html}
	{http://insti.physics.sunysb.edu/\noexpand~siegel/plan.html}}}

%%%%%%%%%%%%%%%%%%%%%%%%%%%%%%%%%%%%%%%%

\pageno=0

\paper

{\rgb{1 0.3 0}{Spacecone quantization\\ \vskip-.4in of AdS superparticle}}

\author\WS\YITP

10-19

%\today
May 27, 2010

We first-quantize the superparticle describing free 10D IIB supergravity on AdS$_5ð$S$^5$.  We choose the worldline coordinate to be a combination of the bulk (spatial) coordinates  of anti de Sitter space and the sphere.  The Hamiltonian is ÓindependentÕ of this ``time" and the fermions.  On the boundary, the representation of PSU(2,2|4) becomes that of projective superspace.  The prepotential propagator reproduces the known field-strength one.

\pageno=2

ÜIntroduction

The Green-Schwarz-style action for the superstring on AdS$_5ð$S$^5$ [1] is complicated by its nonlinearity, particularly in fermions.  In this paper we attack the simpler problem of the superparticle on the same space, which describes the superstring's ground state, 10D IIB supergravity.  We choose a lightcone-type gauge that differs from the usual ones [2] in two ways:  

\item{(1)} The lightcone ``time" is (complex) spatial [3], and comprised of ``bulk" coordinates only, so as to manifestly preserve the SO(3,1) Lorentz and SO(4) internal symmetries of the boundary.  It is also chosen to make the Hamiltonian time-independent. 

\item{(2)} The division of second-class fermionic constraints is such that the imposed constraints eliminate all fermions from the Hamiltonian.  The surviving fermions thus appear trivially in the action, making the propagator (of the field/wave function) a $¶$-function in them.

The representation of the PSU(2,2|4) symmetry generators, after solving the constraints and applying the corresponding gauge conditions, reduces to that of projective superspace (modified by trivial dependence on the ninth dimension [4]), plus time-dependent terms.  The latter can also be found by applying time-dependence in the usual way through the Hamiltonian.

ÜLightcone

The general procedure for choosing lightcone gauges is to (1) determine everything in the action with a(n upper) minus (lightlike) index by the constraints, which are quadratic in covariant derivatives, and (2) gauge fix everything with a plus (canonically conjugate to the previous) using the gauge invariances generated by the same constraints, thus leaving only transverse degrees of freedom.  In flat space, manifest Lorentz invariance is reduced from SO(D$-$1,1) to SO(D$-$2).

In the case of the superparticle, we can use the Casalbuoni-Freund-Brink-Schwarz action [5] without loss of generality, since all formulations lead to the same lightcone, where only physical degrees of freedom survive.  The first-class constraints are
$$ p^2 = 0âÜâp^- = ...,âx^+ =   $$
$$ ©^+ Öpd = 0âÜâ©^- d = ...,â©^+ Ï = 0 $$
for which the quantity multiplying $p^+$ (which is assumed invertible) in the constraint is solved, and we have separated out the independent half of the $û$ constraints.  

This still leaves the remaining second-class constraints
$$ ©^+d = 0,ââÓ©^+ d,©^+ dÕ ¾ ©^+ p^+ $$
half of which can be separated according to some U(1) charge (or discrete symmetry) as first-class.  This half is then set to vanish, along with the conjugate half of $©^- Ï$.  In flat space, the U(1) is chosen as one of the transverse (leaving $p^+$ invariant) Lorentz (rotation) SO(2) generators, further reducing the manifest part of Lorentz invariance to SO(D$-$4)$°$SO(2).

ÜSpacecone

For the AdS case it's useful to start with the parametrization of a group element of PSU(2,2|4) as [6] (in the conventions of [4], which also contains a review)
\vskip-.1in
$$ g = 
\pmatrix{ I & w \cr 0 & I \cr}\pmatrix{ u & 0 \cr 0 & Ðu{}^{-1} \cr}\pmatrix{I & 0 \cr -v & I\cr}
= \pmatrix{ u - wÐu{}^{-1}v & wÐu{}^{-1} \cr -Ðu{}^{-1}v & Ðu{}^{-1} \cr} $$

\noindent where the symmetry group acts on the left and the gauge group on the right.  The derivative form of the generators of these groups is then
\vskip-.2in
$$ G = g»_g = \pmatrix{ w»_w +u»_u & -w»_w w -u»_u w -w»_{Ðu}Ðu -u»_v Ðu \cr
	»_w & -»_w w -»_{Ðu}Ðu \cr} 
	­ \pmatrix{ G_u & -G_v \cr G_w & -G_{Ðu} \cr} $$
$$ D = »_g g = \pmatrix{ »_u u +»_v v & -»_v \cr
	Ðu »_w u +v»_u u +Ðu »_{Ðu} v +v»_v v & -Ðu »_{Ðu} -v»_v \cr} 
	­ \pmatrix{ D_u & -D_v \cr D_w & -D_{Ðu} \cr} $$

\noindent (with ``normal ordering" understood, so all derivatives act only on objects to the right of the generators).

We define $p^+$ by
$$ D_u ® D_{Ðu} ® p^+ I $$
(for identity matrix ``$I$"), which in turn defines how constraints are solved.  (For example, pick only the half of the $û$ constraint that has $p^+$ in it.  Note the sign in the definition of $D_{Ðu}$; the identity piece with opposite relative sign is the ``S" part of PSU.)  This effectively defines $x^-$ as
$$ sdet¼u = sdet¼Ðu = e^{x^-} $$

We then define $x^+$ so that the boundary limit $x^+£0$ scales the two vertical halves of $g$ (i.e., on the gauge group side) oppositely in the above decomposition.  (Scaling them the same would be the ``P" part of PSU.)  This corresponds to 
$$ v¾å{x^+};âu,Ðu¾(x^+)^{1/4};âw¾1 $$
(The boundary limit is the contraction of the gauge groups USp(4)$£$I[USp(2)$^2$], or SO(5)$£$ISO(4).  In terms of the AdS/S radius, this is $x^+£R^2 x^+$, $R£0$.  A perturbation expansion in $R$ is also suggested by a random-lattice approach [7].)
  
ÜQuantization

Besides the constraints/gauge invariances of flat space (Klein-Gordon equation and $û$-symmetry), quadratic in covariant derivatives, we also have the linear ones USp(2,2)$°$USp(4) and ``PS" (of PSU).  We use all these, and a first-class half of the second-class constraints, to fix
$$ \def\normalbaselines{\advance\baselineskip1\jot}
	\matrix{\underline{constraint} & \underline{constrain} & \underline{gauge¼away} \cr
	USp's & p_v,¼some¼p_u & x_v,¼some¼x_u \cr
	PS & some¼p_u & some¼x_u \cr
	p^2 & p^- & x^+ \cr
	(Öpd)_u & d_u & Ï_u  \cr
	d_v & d_v & Ï_v \cr} $$
The quadratic constraints are parts of the matrix square $D^2$ [8].  For abbreviation in this table, we have used ``$x$" for all bosons and ``$Ï$" for all fermions, and ``$p$" and ``$d$" for the corresponding covariant derivatives, with subscripts indicating whether they come from $v$ or $u$ (and $Ðu$).  
The $p_u$ that are fixed (except $p^-$; $p^+$ is not fixed) and all the $d_v$ are fixed identically to 0; the other constrained covariant derivatives are determined in terms of the unconstrained ones, $D_w$ and $p^+$.  Explicitly,
$$ p_v = p_w ¾ å{x^+} »_w,ââp^- ¾{p_w^2\over p^+} ¾ x^+ {»_x^2\over »_-},ââ
	d_u ¾ {p_w d_w\over p^+} ¾ x^+ {»_x »_Ï\over »_-} $$
where we have indicated the $x^+$ (but not $x^-$) dependence.

All the coordinates that survive are $w$ and $x^-$.  (There is also $x^+$, but it can be gauged to 0 in the Schr¬odinger picture, or fixed to $ $ in first-quantization.)  These are exactly the coordinates of 4D N=4 projective superspace, plus the ninth dimension $x^-$, whose momentum $p^+$ counts the number of supergluons in the corresponding conformal field theory on the boundary [4], and thus should be invertible.  These coordinates appear as (using Poincar«e coordinates $(x,x_0)$ and $(y,y_0)$ for both AdS$_5$ and Wick-rotated S$^5$)
$$ w = \pmatrix{ y & ÐÏ \cr Ï & x \cr},ââu = Ðu = \pmatrix{ å{y_0}ÊI & 0 \cr 0 & å{x_0}ÊI \cr}$$
$$ x^+ = x_0 y_0,âx^- = ln\left({y_0\over x_0}\right)âÜâp^+ ¾ »_-,âp^- ¾ x^+ »_+ $$
(A similar choice of $ $ as a combination of an internal coordinate and a physical timelike coordinate has been applied to the pp-wave limit and its lightcone quantization [9].)  Thus the Hamiltonian $H$ appearing in the equation $i»_+=H$ is independent of the ``time" $x^+$.  In these coordinates the metric takes the form [4]
$$ ds^2 = {e^{-x^-}dy^2 - e^{x^-}dx^2 +dx^+ dx^-\over x^+} $$

Since the symmetry generators are $x^+$-scale invariant (homogeneous of degree 0 in $x^+$) before imposing constraints, the only $x^+$ (or $ $) dependence in them comes from solving the constraints, which mix the different sectors.  The result is that these generators take the form of the projective superspace ones (including $p^+$ appearing as a charge), plus terms proportional to $x^+$.  (Projective superspace is defined by replacing all the constraints with simply $D_v=0$, $D_u=D_{Ðu}=»_- I$.)  Thus, in the Schr¬odinger picture the symmety generators are identical to the projective ones, while in the Heisenberg picture they have linear time dependence, as can be generated by $e^{-ix^+ H}$.

ÜPropagator

Since the Hamiltonian is independent of the fermions in the projective representation (i.e., the gauge $Ï_u=Ï_v=0$), the propagator immediately follows from the bosonic result [10]:
$$ ÒV(1) V(2)Ô ¾ ¶^8 (Ï_1-Ï_2) (s_x -s_y)^{-4} $$
$$ s_x = {(x_1-x_2)^2 +(x_{10}-x_{20})^2\over 2x_{10}x_{20}},ââ
	s_y = {(y_1-y_2)^2 +(y_{10}-y_{20})^2\over 2y_{10}y_{20}} $$

For comparison, we now rederive the propagator for the chiral field strength.  The difference in division of second-class constraints (``chirality"), now distinguishing SL(2,C) dotted (barred) and undotted spinors, is
$$ d_v V = Ðd_v V = 0;ââd_v  = Ðd_w  = 0,âÐd_v Ѝ = d_w Ѝ = 0 $$
where
$$ Ód_v,d_wÕ ¾ ÓÐd_v,Ðd_wÕ ¾ p^+,ââÓd_v,Ðd_wÕ = ÓÐd_v,d_wÕ = 0 $$
$$ Ód_v,d_vÕ = Ód_v,Ðd_vÕ = Ód_w,d_wÕ = Ód_w,Ðd_wÕ = 0 $$
(10D ``chirality" means depending only on the lightcone 8-spinor of positive U(1) charge, namely $Ï_w ¢ ÐÏ_v$, in a lightcone representation, where the other half of the 16-spinor, $Ï_u¢ÐÏ_{Ðu}$, is again gauged away by $û$ symmetry.)

The solution to the reality condition on the chiral field strength is then given in terms of the prepotential by
$$ d_w^4  = Ðd_w^4 ЍâÜ⍠= Ðd_w^4 V,âЍ = d_w^4 V $$
(There are also redundant reality conditions, $Ðd_v^4  = d_v^4 Ѝ$ and others from switching various numbers of $d_w$ with $Ðd_v$.)  This is essentially a Fourier transform in the fermions (up to powers of $p^+$), replacing $ÐÏ_w$'s in $V$ with $ÐÏ_v$'s in $$.
The field-strength propagator is then
$$ ҍ(1) Ѝ(2)Ô = Ðd_{1w}^4 d_{2w}^4 ÒV(1) V(2)Ô ¾ (ös_x -ös_y)^{-4} $$
where $ös_x$ and $ös_y$ are the chiral-antichiral versions of $s_x$ and $s_y$ [11] (by adding fermion terms):
$$ d_{1v}ös_{x,y} = Ðd_{1w}ös_{x,y} = Ðd_{2v}ös_{x,y} = d_{2w}ös_{x,y} = 0 $$
(Chiral superspace corresponds to the ``complex gauge" considered in [12].  It seems unlikely that a lightcone-type gauge based on chiral superfields could give a manifestly SO(3,1) covariant formulation on the boundary, although it might relate to the 4D lightcone one.)

An analog of this construction is the corresponding propagator for 4D N=1 supersymmetry, where a chiral superfield can also be written in terms of a real prepotential [13] (although this is unnecessary for the construction).  In an appropriate gauge,
$$ ÒÄ(1) ÐÄ(2)Ô = Ðd_1^2 d_2^2 ÒV(1) V(2)Ô ¾ Ðd_1^2 d_2^2 ¶^4(Ï_1-Ï_2)x^{-2} 
	= (x+iÏ_1 ÐÏ_2)^{-2} $$

ÜAcknowledgments

I thank Machiko Hatsuda for helpful discussions.  This work is supported in part by National Science Foundation Grant No.¼PHY-0653342.

\refs

£1 %\cite{Metsaev:1998it}
%\bibitem{Metsaev:1998it}
  R.R. Metsaev and A.A. Tseytlin,
  %``Type IIB superstring action in AdS(5) x S(5) background,''
  ÓNucl. Phys.  BÕ {\bf 533} (1998) 109
  \xxxlink{hep-th/9805028}.
  %%CITATION = NUPHA,B533,109;%%

£2 %\cite{Metsaev:1999gz}
%\bibitem{Metsaev:1999gz}
  R.R. Metsaev,
  %``Light cone gauge formulation of IIB supergravity in AdS(5) x S(5)
  %background and AdS/CFT correspondence,''
  ÓPhys. Lett.  BÕ {\bf 468} (1999) 65
  \xxxlink{hep-th/9908114};
  %%CITATION = PHLTA,B468,65;%%
\\
%\cite{Pesando:1999wf}
%\bibitem{Pesando:1999wf}
  I. Pesando,
  %``On the fixing of the kappa gauge symmetry on AdS and flat background:  The
  %lightcone action for the type IIb string on AdS(5) x S(5),''
  ÓPhys. Lett.  BÕ {\bf 485} (2000) 246
  \xxxlink{hep-th/9912284};
  %%CITATION = PHLTA,B485,246;%%
\\
%\cite{Metsaev:2000yf}
%\bibitem{Metsaev:2000yf}
  R.R. Metsaev and A.A. Tseytlin,
  %``Superstring action in AdS(5) x S(5): kappa-symmetry light cone gauge,''
  ÓPhys. Rev.  DÕ {\bf 63} (2001) 046002\\
  \xxxlink{hep-th/0007036};
  %%CITATION = PHRVA,D63,046002;%%
\\
%\cite{Metsaev:2000yu}
%\bibitem{Metsaev:2000yu}
  R.R. Metsaev, C.B. Thorn and A.A. Tseytlin,
  %``Light-cone superstring in AdS space-time,''
  ÓNucl. Phys.  BÕ {\bf 596} (2001) 151\\
  \xxxlink{hep-th/0009171}.
  %%CITATION = NUPHA,B596,151;%%

£3 %\cite{Chalmers:1998jb}
%\bibitem{Chalmers:1998jb}
  G. Chalmers and W. Siegel,
  %``Simplifying algebra in Feynman graphs. II: Spinor helicity from the
  %spacecone,''
  ÓPhys. Rev.  DÕ {\bf 59} (1999) 045013
  \xxxlink{hep-ph/9801220}.
  %%CITATION = PHRVA,D59,045013;%%

£4 %\cite{Siegel:2010yd}
%\bibitem{Siegel:2010yd}
  W. Siegel,
  %``AdS/CFT in superspace,''
 \xxxlink{1005.2317} [hep-th].
  %%CITATION = ARXIV:1005.2317;%%

£5 R. Casalbuoni, ÓPhys. Lett.Õ É62B (1976) 49;\\
P.G.O. Freund, unpublished, cited ã
%\cite{Ferber:1977qx}
%\bibitem{Ferber:1977qx}
  A. Ferber,
  %``Supertwistors And Conformal Supersymmetry,''
  ÓNucl. Phys.  BÕ {\bf 132} (1978) 55;
  %%CITATION = NUPHA,B132,55;%%
\\
L. Brink and J.H. Schwarz, ÓPhys. Lett.Õ É100B (1981) 310.

£6 %\cite{Hatsuda:2002wf}
%\bibitem{Hatsuda:2002wf}
  M. Hatsuda and W. Siegel,
  %``A new holographic limit of AdS(5) x S**5,''
  ÓPhys. Rev.  DÕ {\bf 67} (2003) 066005
  \xxxlink{hep-th/0211184},
  %%CITATION = PHRVA,D67,066005;%%
\\
%\cite{Hatsuda:2007wr}
%\bibitem{Hatsuda:2007wr}
%  M.~Hatsuda and W.~Siegel,
  %``Superconformal spaces and implications for superstrings,''
  ÓPhys. Rev.  DÕ {\bf 77} (2008) 065017
  \xxxlink{0709.4605} [hep-th].
  %%CITATION = PHRVA,D77,065017;%%

£7 %\cite{Nastase:2000za}
%\bibitem{Nastase:2000za}
  H. Nastase and W. Siegel,
  %``A new AdS/CFT correspondence,''
  ÓJHEPÕ {\bf 0010} (2000) 040
  \xxxlink{hep-th/0010106}.
  %%CITATION = JHEPA,0010,040;%%

£8 %\cite{Hatsuda:2001xf}
%\bibitem{Hatsuda:2001xf}
  M. Hatsuda and K. Kamimura,
  %``Classical AdS superstring mechanics,''
  ÓNucl.\ Phys.\  B Õ{\bf 611} (2001) 77
  \xxxlink{hep-th/0106202}.
  %%CITATION = NUPHA,B611,77;%%

£9 %\cite{Blau:2002dy}
%\bibitem{Blau:2002dy}
  M. Blau, J.M. Figueroa-O'Farrill, C. Hull and G. Papadopoulos,
  %``Penrose limits and maximal supersymmetry,''
  ÓClass. Quant. Grav. Õ {\bf 19} (2002) L87
  \xxxlink{hep-th/0201081};\\
  %%CITATION = CQGRD,19,L87;%%
%\cite{Berenstein:2002jq}
%\bibitem{Berenstein:2002jq}
  D.E. Berenstein, J.M. Maldacena and H.S. Nastase,
  %``Strings in flat space and pp waves from N = 4 super Yang Mills,''
  ÓJHEPÕ {\bf 0204} (2002) 013\\
  \xxxlink{hep-th/0202021};\\
  %%CITATION = JHEPA,0204,013;%%
%\cite{Hatsuda:2002xp}
%\bibitem{Hatsuda:2002xp}
  M. Hatsuda, K. Kamimura and M. Sakaguchi,
  %``From super-AdS(5) x S**5 algebra to super-pp-wave algebra,''
  ÓNucl. Phys.  BÕ {\bf 632} (2002) 114\\
  \xxxlink{hep-th/0202190}.
  %%CITATION = NUPHA,B632,114;%%

£10 %\cite{Dorn:2003au}
%\bibitem{Dorn:2003au}
  H. Dorn, M. Salizzoni and C. Sieg,
  %``On the propagator of a scalar field on AdS x S and on the BMN plane
  %wave,''
  ÓJHEPÕ {\bf 0502} (2005) 047
  \xxxlink{hep-th/0307229}.
  %%CITATION = JHEPA,0502,047;%%

£11 %\cite{Dai:2009zg}
%\bibitem{Dai:2009zg}
  P. Dai, R.-N. Huang and W. Siegel,
  %``Covariant propagator in AdS5 x S5 superspace,''
  ÓJHEPÕ {\bf 1003} (2010) 001
  \xxxlink{0911.2211} [hep-th].
  %%CITATION = JHEPA,1003,001;%%

£12 %\cite{Roiban:2000yy}
%\bibitem{Roiban:2000yy}
  R. Roiban and W. Siegel,
  %``Superstrings on AdS(5) x S(5) supertwistor space,''
  ÓJHEPÕ {\bf 0011} (2000) 024
  \xxxlink{hep-th/0010104}.
  %%CITATION = JHEPA,0011,024;%%

£13 %\cite{Gates:1980ay}
%\bibitem{Gates:1980ay}
  S.J. Gates, Jr.,
  %``Super P Form Gauge Superfields,''
  ÓNucl. Phys.  BÕ {\bf 184} (1981) 381.
  %%CITATION = NUPHA,B184,381;%%

\bye